\documentclass[12pt,a4paper]{article}
\usepackage[T2A]{fontenc}
\usepackage{amsfonts}
\usepackage{amsmath}
\usepackage{amssymb}
\usepackage[english]{babel}
\usepackage{graphicx}

\begin{document}
\title{Discrete mechanics: a sequential growth dynamics for causal sets that is based on binary alternatives}
\author{Alexey L. Krugly\thanks{Scientific Research Institute for System Analysis of the Russian Academy of Science, 117218, Nahimovskiy pr., 36, k. 1, Moscow, Russia; akrugly@mail.ru.}}
\date{} \maketitle
\begin{abstract}
One of approaches to quantum gravity is different models of a discrete pregeometry. An example of a discrete pregeometry on a microscopic scale is introduced. This is the particular case of a causal set. The causal set is a locally finite partially ordered set. The dynamics of this model is a stochastic sequential growth dynamics. New elements of causal set are added one by one. The probability of this addition of a new element depends on the structure of existed causal set. The particular case of the dynamics is considered. This dynamics is based on binary alternatives. Each directed path is considered as a sequence of outcomes of binary alternatives. The probabilities of a stochastic sequential growth have quadratic dependence on these paths.
\end{abstract}
\newpage
\tableofcontents
\newpage
\section{INTRODUCTION\label{I}}

We adopt the extreme viewpoint provided by causal set program (see e.g. \cite{Sorkin2005,Dowker2006,Henson2009,Wallden2010}, an approach to quantum gravity pioneered by J. Myrheim \cite{Myrheim} and G. 't Hooft \cite{'t Hooft}. The usual spacetime continuum is thus regarded as a macroscopic construct that does not exist. On a microscopic scale spacetime is a partially ordered set of purely discrete points.

In mathematical terms, a causal set is a pair ($\mathcal{C}$, $\prec$), where $\mathcal{C}$ is a set and $\prec$ is a binary relation on $\mathcal{C}$ satisfying the following properties ($a,\ b,\ c$ are general points in $\mathcal{C}$):
\begin{equation}
\label{eq:I1.2} a\prec a\qquad\textrm{(irreflexivity),}
\end{equation}
\begin{equation}
\label{eq:I1.3} \{a\mid(a\prec b)\wedge(b\prec a)\}=\emptyset \qquad \textrm{(acyclicity),}
\end{equation}
\begin{equation}
\label{eq:I1.4} (a\prec b)\wedge(b\prec c)\Rightarrow(a\prec c)\qquad \textrm{(transitivity),}
\end{equation}
\begin{equation}
\label{eq:I1.5} \mid\mathcal{A}(a, b)\mid<\infty\qquad\textrm{(local finiteness),}
\end{equation}
where $\mathcal{A}(a,\ b)$ is an Alexandrov set of the elements $a$ è $b$. $\mathcal{A}(a,\ b)=\{c\mid a\prec c \prec b\}$. The local finiteness means that the Alexandrov set of any elements is finite. Sets of elements are denoted by calligraphic capital Latin letters.

The main hypothesis of this study is that the causal set must describe the matter and its elements are primitive material objects. Any complicated object is a structure that consists of the elements of causal set. Spacetime continuum emerges only on a macroscopic level and we must describe particles without any reference to spacetime as some repetitive structures \cite{1004.5077}. According to this approach, the world is a collection of cyclic structures. One cyclic structure can include other cyclic structures and there is the hierarchy of cyclic structures. These structures must emerge as the result of some dynamical self-organization. The example of such dynamics is introduced in \cite{1004.5077}. Our goal is a consideration of another example of such dynamics that is based on binary alternatives.

\section{THE MODEL\label{M}}

\begin{figure}
	\centering	
		\includegraphics[trim=8cm 18cm 8cm 9cm]{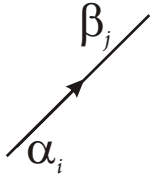}
	\caption{A chronon.}
	\label{fig:fig1}
\end{figure}

\begin{figure}
	\centering	
		\includegraphics[trim=8cm 18cm 8cm 7cm]{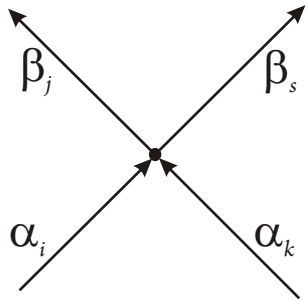}
	\caption{An x-structure.}
	\label{fig:fig2}
\end{figure}

The central hypothesis is that a physical process is a finite network of finite elementary processes \cite{STC, STC2, STC3, STC4}. The primitive entities of the physical world have not any internal structure. This is primordial indivisible objects. Consequently, they itself have not any internal properties except one. They exist. The property ``existence'' can adopt two values: ``the primitive entity exists'', and ``the primitive entity does not exist''. The primitive entity is called a material point. The primitive process can be thought of as act of creation. The value of the property ``existence'' of the material point varies from ``the material point does not exist'' to ``the material point exists'' by this process. Its dual represents the act of destruction. These primitive processes are called monads \cite{STC4}. A propagation of the material point is simply an ordered pair of creation and annihilation. This process of propagation is called a chronon \cite{STC}. A chronon can be represented by a diagram as an arrow or a directed edge (Fig.\ \ref{fig:fig1}). The meaning of letters in the figure will be explained below.
The general process will be a collection of creations and annihilations. The material point can be destroyed only by the interaction with another material point. The interaction of this second material point means the change of its state. Only one kind of change is possible. This is the annihilation of the second material point. Suppose the number of the material points does not change. This is a fundamental conservation law. We have a simplest interaction process: two material points are destroyed and two material points are created. This process is called a tetrad \cite{FinkMcC1} or an x-structure \cite{Krugly2002} An x-structure can be represented by a diagram as a vertex with incident edges (Fig.\ \ref{fig:fig2}).
The x-structure consists of two monads of destruction and two monads of creation. Suppose any process can be divided into x-structures. This symmetric dyadic kinematics is called for brevity X kinematics \cite{Fink88}. A chronon and an x-structure describe an immediate causal priority of monads. If a creation causally precedes an annihilation this is a chronon. If an annihilation causally precedes a creation this is an x-structure. Suppose there is a universal causal order of monads.

In the considered model any process can be described as some graph. But such description is not useful. By definition, a graph is a set of vertexes and a binary relation (edges) over this set. We cannot describe external lines as in Feynman diagrams. For example, an x-structure is not a graph. We can define a set of edges, and vertexes as a relation over this set. But in this case, if we divide a structure into substructures, we must duplicate the edges which connect these substructures. This is not useful either. It is convenient to break the edge into two halves, monads, of which the edge is regarded as composed \cite{FinkMcC1}.

Consider the axiomatic approach to this model. Consider the set $\mathcal{G}$ of monads and a binary relation (an immediate causal priority) over this set. By $\alpha_i$ and $\beta_j$ denote the monad of creation and destruction, respectively. By $(\alpha_i\beta_j)$ denote an immediate causal priority relation of $\alpha_i$ and $\beta_j$. $\mathcal{G}$ satisfies the following axioms.
\begin{equation}
\label{eq:M1.1}
\forall\alpha_i(\exists !\beta_j(\alpha_i\beta_j))\vee(\not\exists\beta_j(\alpha_i\beta_j))\textrm{,}
\end{equation}
\begin{equation}
\label{eq:M1.2}
\forall\alpha_i(\not\exists \alpha_j(\alpha_i\alpha_j))\textrm{,}
\end{equation}
\begin{equation}
\label{eq:M1.3}
\forall\beta_j(\exists !\alpha_i(\alpha_i\beta_j))\vee(\not\exists\alpha_i(\alpha_i\beta_j))\textrm{,}
\end{equation}
\begin{equation}
\label{eq:M1.4}
\forall\beta_j(\not\exists \beta_i(\beta_i\beta_j))\textrm{.}
\end{equation}
There is no more than one monad $\beta_j$ and does not exist the monad $\alpha _j$ which immediately causally follow $\alpha_i$. There is no more than one monad $\alpha_i$ and does not exist the monad $\beta_i$ which immediately causally precede any $\beta_j$. The pair $(\alpha_i\beta_j)$ is called a chronon or an edge (Fig.\ \ref{fig:fig1}).

The following axioms describe an x-structure (Fig.\ \ref{fig:fig2}).
\begin{equation}
\label{eq:M1.5}
\forall\alpha_i\exists !\alpha_j(\forall\beta_k(\beta_k\alpha_i)\Rightarrow(\beta_k\alpha_j))\textrm{,}
\end{equation}
\begin{equation}
\label{eq:M1.6}
\forall\beta_i\exists !\beta_j(\forall\alpha_k(\beta_i\alpha_k)\Rightarrow(\beta_j\alpha_k))\textrm{.}
\end{equation}
There is two and only two monads $\beta_i$ and $\beta_j$ which immediately causally precede any $\alpha_k$. There is two and only two monads $\alpha_i$ and $\alpha_j$ which immediately causally follow any $\beta_k$.

A causality is described by the following axiom.
\begin{equation}
\label{eq:M1.7}
\{\alpha_i|(\alpha_i\beta_k)(\beta_k\alpha_l)\dots(\beta_j\alpha_i)\}= \emptyset\textrm{.}
\end{equation}
We consider only finite sets of monads.
\begin{equation}
\label{eq:M1.8} |\mathcal{G}|<\infty\textrm{.}
\end{equation}

$\mathcal{G}$ is called a dynamical graph or a d-graph. The properties of d-graph are described in \cite{1008.5169}. Some definitions and properties will be useful for the following work.

The monad of any type is denoted by $\gamma_i$. The monad $\gamma_i$ may be $\alpha_i$ or $\beta_i$. Any monad belongs to an x-structure. Any d-graph can be formed from the empty set of monads by sequential adding of x-structures one after another.

Define the isomorphism that takes each x-structure to a vertex of some graph and each chronon to an edge of this graph. We get the directed acyclic graph. All properties of this graph are the properties of the d-graph. For this reason the considered set of monads is called a d-graph. This isomorphism is used in the figures for simplicity. The edges are figured without a partition into monads. The monads are figured by placing $\alpha_i$ and $\beta_j$ near the beginning and the end of the edge, respectively, as necessary. The x-structures are figured by big black points.

A sequence of monads is a saturated chain or a path if each monad immediately causally precedes the sequent monad. Two monads are causally connected if they are connected by the path. The causal connection is denoted by $\prec$. The first monad of the path is called a cause. The last monad of the path is called an effect. Two monads $\gamma_i$ and $\gamma_j$ are causally unconnected iff neither $\gamma_i\prec\gamma_j$ nor $\gamma_j\prec\gamma_i$.

A d-graph is a causal set of monads, and a causal set of chronons, and a causal set of x-structures. The subset of monads $\alpha_i$ and the subset of monads $\beta_j$ are causal sets.

The past of the monad is the set of monads, which causally precede this monad. The past of $\gamma_i$ is denoted by $\mathcal{P}(\gamma_i)=\{\gamma_j|\gamma_j\prec\gamma_i\}$. The future of the monad is the set of monads, which causally follow this monad. The future of $\gamma_i$ is denoted by $\mathcal{F}(\gamma_i)=\{\gamma_j|\gamma_j\succ\gamma_i\}$.

A monad is called maximal iff its future is an empty set. Any maximal monad is a monad of a type $\alpha$. A monad is called minimal iff its past is an empty set. Any minimal monad is a monad of a type $\beta$. Maximal and minimal monads are called external monads. Other monads are called internal monads. The monad is internal iff it is included in a chronon.

A chain is a totally (or a linearly) ordered subset of monads. Every two monads of this subset are related by $\prec$. A chain is a subset of a path.

An antichain is a totally unordered subset of monads. Every two elements of this subset are not related by $\prec$. The cardinality of an antichain is called a width of an antichain.

A slice is a maximal antichain. Every monad in $\mathcal{G}$ is either in the slice or causal connected to one of its monads. The set of all maximal (or minimal) monads is a slice. A slice is a discrete spacelike hypersurface.

All slices of $\mathcal{G}$ have the same width. This width is called a width of $\mathcal{G}$. This is the conservation law of the number of material points. The number of material points remains in each elementary interaction (x-structure). Consequently the number of material points remains in each process.

\section{THE SEQUENTIAL GROWTH DYNAMICS\label{SGD}}

Consider the following concept of a d-graph dynamics. The past and the future exist, are determined, and are changeless. In the discrete mechanics this means that the infinite d-graph of the universe exists. Any d-graph is a d-subgraph of the infinite d-graph of the universe. It is meaningless to talk about the exact structure of the d-graph if we cannot determine its structure. The structure of the infinite d-graph of the universe implies the infinite amount of information. But we can only actually know a finite number of facts. Therefore any observer can consider only finite fragments and take into account the rest of the universe in an approximate way.

We have the following assumption. Any d-graph has the certain structure. We can determine the structure of any d-graph.

Suppose we have the information about the structure of some d-graph $\mathcal{G}$. This is the description of some part of some physical process. The task is to predict the future stages of this process or to reconstruct the past stages. This mean to determine the structure of the d-subgraph that is connected with $\mathcal{G}$. In general case, we cannot determine this structure unambiguously. We can only calculate probabilities of different variants. In particular case, the probability of some variant can be equal to 1. This is a deterministic process.

The aim of the d-graph dynamics is to calculate the probability of each variant of the structure of the d-graph $\mathcal{\tilde {G}}$ that is connected with the given $\mathcal{G}$. We can reconstruct the structure of $\mathcal{\tilde {G}}$ step by step. The minimal part is an x-structure. We start from some given d-graph $\mathcal{G}$ and add new x-structures to $\mathcal{G}$ one by one. This procedure is proposed in \cite{ Krugly2002,Krugly1998}. Similar procedure and the term `a classical sequential growth dynamics' is proposed in \cite{RideoutSorkin} for another model of a causal set. The addition of one x-structure is called an elementary extension.

We can determine the structure of the d-graph after each elementary extension by assumption. This is not an appearance of new parts of the infinite d-graph of the universe. This is an appearance of new information about the existing infinite d-graph of the universe. We can randomly initiate one elementary extension for any given d-graph and we can determine the exact change of the structure of the d-graph that is the result of this elementary extension. This procedure is called the elementary measurement. The sequence of elementary measurements is the sequence of the obtaining of the information about the structure of the d-graph by the observer.

In quantum theory, a set of results of sequential measurements is a classical stochastic sequence. Similarly, a sequence of elementary extensions is a classical stochastic sequence. We have two different times. One time is the casual order of a d-graph. This is the time of the object. Another time is the sequence of elementary extensions. This is the time of the observer.

There are different ways to add x-structure to the same d-graph. Consider the types of elementary extensions.

First type is an internal elementary extension to the future (Fig.\ \ref{fig:fig3}). In this and following figures the d-graph $\mathcal{G}$ is represented by a rectangle because it can have an arbitrary structure. The external monads are represented by arrows. The arrows of minimal monads are directed to the rectangle. The arrows of maximal monads are directed from the rectangle. This elementary extension describes the future evolution of the process without any interaction with environment. The width $n$ of $\mathcal{G}$ is not changed by this elementary extension.
\begin{figure}
	\centering	
		\includegraphics[width=4cm,trim=8cm 16cm 8cm 5cm]{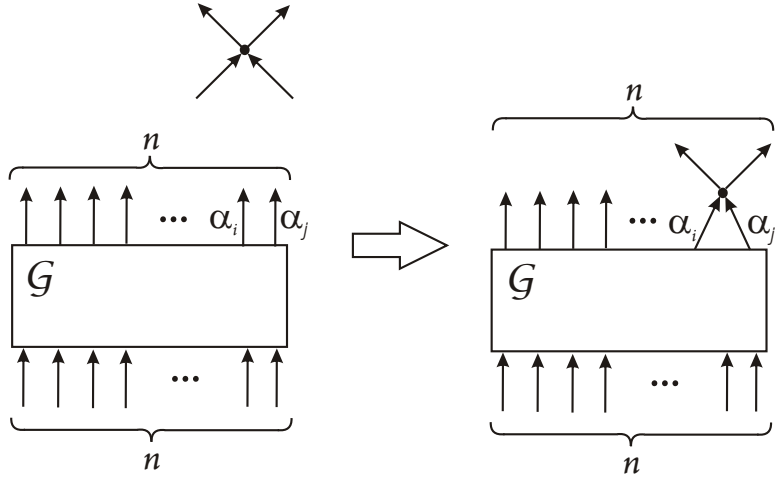}
	\caption{The internal elementary extension to the future.}
	\label{fig:fig3}
\end{figure}

\begin{figure}
	\centering	
		\includegraphics[width=4cm,trim=8cm 16cm 8cm 5cm]{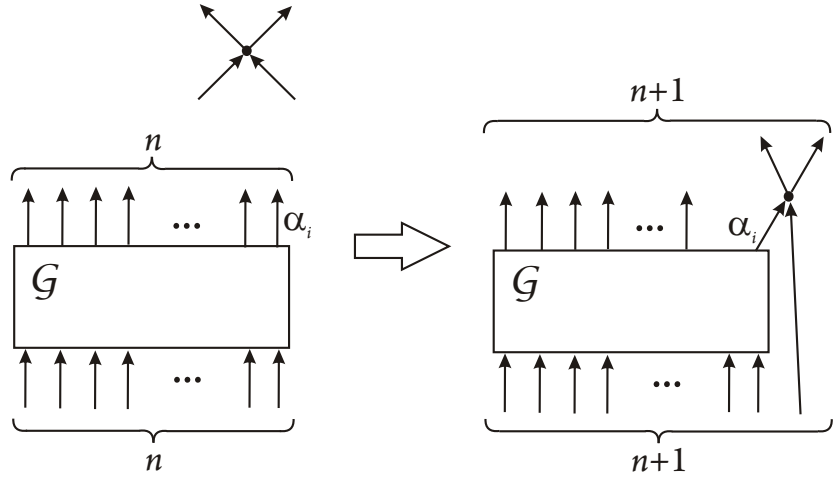}
	\caption{The external elementary extension to the future.}
	\label{fig:fig4}
\end{figure}

Second type is an external elementary extension to the future (Fig.\ \ref{fig:fig4}). This elementary extension describes the future interaction of the process with environment. The width $n$ of $\mathcal{G}$ has increased by 1.

\begin{figure}
	\centering	
		\includegraphics[width=4cm,trim=8cm 14cm 8cm 7cm]{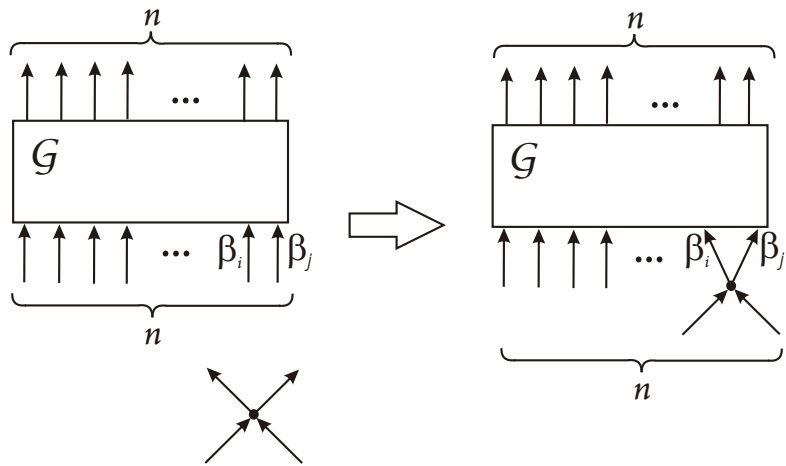}
	\caption{The internal elementary extension to the past.}
	\label{fig:fig5}
\end{figure}

Third type is an internal elementary extension to the past (Fig.\ \ref{fig:fig5}). This elementary extension describes the past evolution of the process without any interaction with environment. The width $n$ of $\mathcal{G}$ is not changed by this elementary extension.

\begin{figure}
	\centering	
		\includegraphics[width=4cm,trim=8cm 14cm 8cm 7cm]{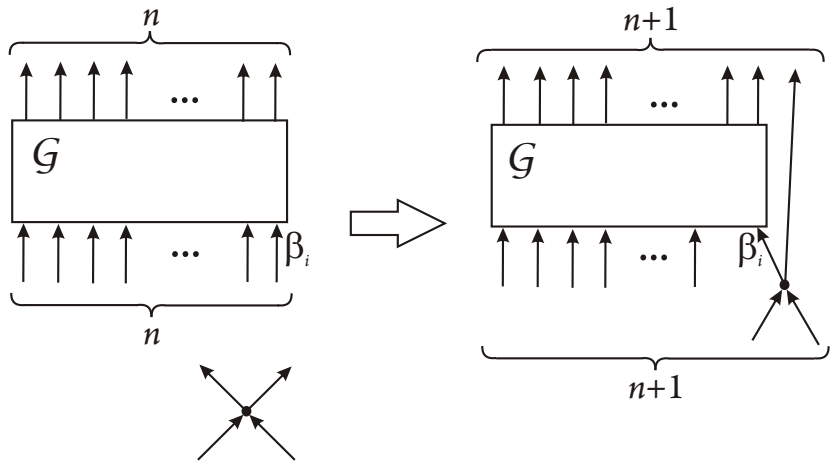}
	\caption{The external elementary extension to the past.}
	\label{fig:fig6}
\end{figure}

Fourth type is an external elementary extension to the past (Fig.\ \ref{fig:fig6}). This elementary extension describes the past interaction of the process with environment. The width $n$ of $\mathcal{G}$ has increased by 1.

We can get every connected d-graph by a sequence of elementary extensions of these types \cite[Teorem~2]{1008.5169}. Other types of elementary extensions are unnecessary.

\section{BINARY ALTERNATIVES \label{BA}}

In each step, we need to calculate the probabilities of the finite set of events. The probabilities can be composed of some primordial entities. Such simplest entity is a binary alternative. \flqq It is certainly possible to decide any large alternative step by step in binary alternatives\frqq \ \cite[p. 222]{Weiz1}. \flqq The use of discrete alternatives is justified by the remark that we can only actually know a finite number of facts\frqq \ \cite[p. 100]{Weiz2}. This alternative has two outcomes with probabilities $1/2\times 1/2$.

Binary alternatives are discussed in \cite[section~44.5]{gr} as background of dynamics on a microscopic scale. The idea of a binary alternative was developed in the theory of ur object (see e. g. \cite{FinkMcC1, Weiz1, Weiz2, FinkMcC2, Weiz3}). This list of references is by no means complete (also see \cite{MK} and references therein). The denotation ur object is derived from the German prefix ur- which means something like original, elementary or primordial. \flqq $\dots$ the decision of an elementary binary alternative is the elementary process and hence the elementary interaction\frqq \ \cite[p. 94]{Weiz2}. This is a process that has 1 bit of information.

In the considered model we can identify a binary alternative with an x-structure. If we choose a directed path from any monad in a d-graph, we must choose one of two outcomes of the alternative in each x-structure. Assume the equal probabilities for both outcomes. Consequently if a directed path passes through $k$ x-structures, this path has the probability $2^{-k}$. We have the same choice for opposite directed path.

Usually it is considered only two values of a causal connection between two events: the causal connection exists or does not exist. Introduce amplitude $a_{ij}$ of causal connection of two monads $\gamma_i$ and $\gamma_j$ in a d-graph. By definition, put
\begin{equation}
\label{eq:BA1.1} a_{ij}=\sum_{m=1}^M 2^{-k(m)}\textrm{,}
\end{equation}
where $M$ is the number of directed paths between $\gamma_i$ and $\gamma_j$ and $k(m)$ is the number of x-structures through which the directed path number $m$ passes. Obviously, $a_{ij}= a_{ji}$. This definition has clear physical meaning. The causal connection of two monads is stronger if there are more directed paths between these monads and these paths are shorter.

\section{THE ALGORITHM FOR CALCULATING PROBABILITIES \label{ACP}}

Consider the d-graph $\mathcal{G}$. The width of $\mathcal{G}$ is equal to $n$. We calculate the probabilities of elementary extensions of $\mathcal{G}$ by three steps.

The first step is the choice of the elementary extension to the future or to the past. We assume time symmetry. Then the probability of this choice is $1/2\times 1/2$. We can consider a time asymmetric model if we choose other probabilities.

A new x-structure is added to one or two external monads. The second step is the equiprobable choice of one external monad that takes part in the elementary extension. This is a maximal monad if we choose the future evolution in the first step. Otherwise this is a minimal monad. The probability of this choice is $1/n$.

\begin{figure}[ht]
	\centering	
		\includegraphics[width=4cm,trim=8cm 16cm 8cm 7cm]{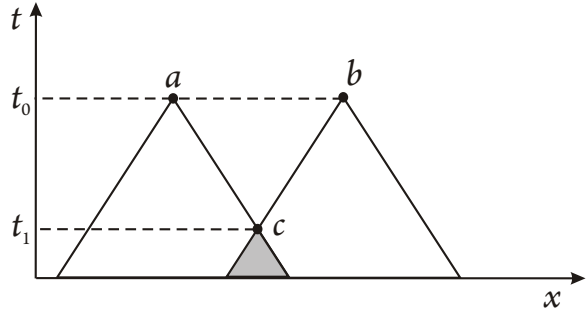}
	\caption{The connection of two events by the common past.}
	\label{fig:fig7}
\end{figure}

The probability of the choice of a second monad which takes part in the elementary extension must depend on the connection between this monad and the first monad which was chosen in the second step. In relativity theory two events $a$ and $b$ cannot be connected in the time instant (Fig.\ \ref{fig:fig7}).
They are connected only by the intersection of their past light cones (the shaded triangle in the figure). This is their common past. Similarly, if we reconstruct the evolution in the past, two events are connected by the common future.

Number the maximal monads by latin indices from 1 to $n$. Number the minimal monads by latin indices with bar from $\bar 1$ to $\bar n$. Describe the common past of the maximal monads and the common future of the minimal monads by the amplitudes of causal connection. By $p_{ij}$ denote the conditional probability of the addition of a new x-structure to the maximal monads $\alpha_i$ and $\alpha_j$ if we chose the maximal monad $\alpha_i$. By definition, put 
\begin{equation}
\label{eq:ACP1.1} p_{ij}=\sum_{\bar s=1}^n a_{i\bar s} a_{\bar sj}\textrm{.}
\end{equation}

\begin{figure}[ht]
	\centering	
		\includegraphics[width=4cm,trim=8cm 15cm 8cm 7cm]{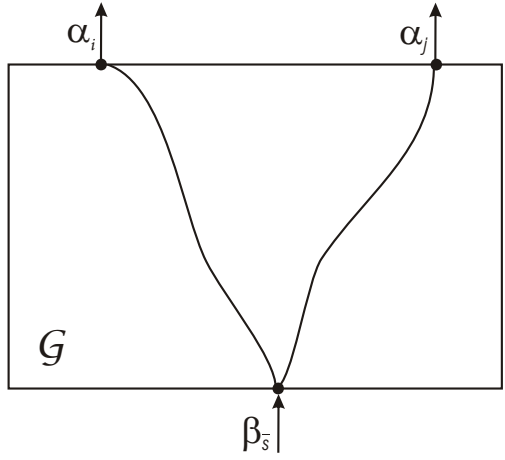}
	\caption{The summable path.}
	\label{fig:fig8}
\end{figure}

In this equation, we sum up all paths from the maximal monad $\alpha_i$ to each minimal monad and then to the maximal monad $\alpha_j$. Each path consists of two parts. The first part is the opposite directed path from the maximal monad $\alpha_i$ to any minimal monad $\beta_{\bar s}$, and the second part is the directed path from the minimal monad $\beta_{\bar s}$ to the maximal monad $\alpha_j$ (Fig.\ \ref{fig:fig8}).
The paths are summed up with coefficients in compliance with (\ref{eq:BA1.1}).

If $i=j$ this equation defines the probability of the addition of a new x-structure to one maximal monad $\alpha_i$. Similarly, define the addition of a new x-structure to the past.
\begin{equation}
\label{eq:ACP1.2} p_{\bar i\bar j}=\sum_{s=1}^n a_{\bar is} a_{s\bar j}\textrm{.}
\end{equation}

These probabilities are normalized to the unity. Each directed path from the minimal monad $\beta_{\bar s}$ ends in some maximal monad. We get
\begin{equation}
\label{eq:ACP1.3} \sum_{j=1}^n a_{\bar sj}=1\textrm{.}
\end{equation}
This sum includes all binary choices. Similarly, each opposite directed path from the maximal monad $\alpha_i$ ends in some minimal monad. We get
\begin{equation}
\label{eq:ACP1.4} \sum_{j=1}^n p_{ij}=\sum_{j=1}^n \sum_{\bar s=1}^n a_{i\bar s} a_{\bar sj}=1\textrm{.}
\end{equation}

The third step is the choice of second external monad with the probability (\ref{eq:ACP1.1}) or (\ref{eq:ACP1.2}). These three steps define the probability of any elementary extension.

We can express the equations (\ref{eq:ACP1.1}) and (\ref{eq:ACP1.2}) in a matrix form. Introduce a matrix $(\mathbf{a})$ of amplitudes. An element $a_{\bar ij}$ of this matrix is equal to the amplitude of causal connection of the minimal monad $\beta_{\bar i}$ and the maximal monad $\alpha_j$. The matrix $(\mathbf{a})$ is a square matrix of size $(n, n)$. Introduce matrixes $(\mathbf{p})_f$ and $(\mathbf{p})_p$. An element number $ij$ of $(\mathbf{p})_f$ is equal to $p_{ij}$. An element number $\bar i\bar j$ of $(\mathbf{p})_p$ is equal to $p_{\bar i\bar j}$. We have
\begin{equation}
\label{eq:ACP1.5} (\mathbf{p})_f= (\mathbf{a})^T(\mathbf{a}) \textrm{,}
\end{equation}
\begin{equation}
\label{eq:ACP1.6} (\mathbf{p})_p= (\mathbf{a})(\mathbf{a})^T \textrm{.}
\end{equation}
The sum of the elements in each row and in each column is equal to 1 for the matrixes $(\mathbf{a})$, $(\mathbf{p})_f$, and $(\mathbf{p})_p$.

Combine all three steps. We have the following equations for each type of elementary extension.

We get for the probability $P_{ij}$ of the elementary extension of first type (Fig.\ \ref{fig:fig3})
\begin{equation}
\label{eq:ACP1.7} P_{ij}= 1/(2n)\sum_{\bar s=1}^n (a_{i\bar s} a_{\bar sj}+a_{j\bar s} a_{\bar si})=1/n\sum_{\bar s=1}^n a_{i\bar s} a_{\bar sj} \textrm{.}
\end{equation}
The two summands correspond to two ways to get the same elementary extension. We can choose the maximal monad $\alpha_i$ with the probability $1/n$, then we can choose the maximal monad $\alpha_j$ with the probability $p_{ij}$. We can also choose the maximal monad $\alpha_j$ with the probability $1/n$, then we can choose the maximal monad $\alpha_i$ with the probability $p_{ji}$.

We get for the probability $P_{ii}$ of the elementary extension of second type (Fig.\ \ref{fig:fig4})
\begin{equation}
\label{eq:ACP1.8} P_{ii}= 1/(2n)\sum_{\bar s=1}^n a_{i\bar s} a_{\bar si} \textrm{.}
\end{equation}

We get for the probability $P_{\bar i\bar j}$ of the elementary extension of third type (Fig.\ \ref{fig:fig5})
\begin{equation}
\label{eq:ACP1.9} P_{\bar i\bar j} = 1/(2n)\sum_{s=1}^n (a_{\bar is} a_{s\bar j}+a_{\bar js} a_{s\bar i})= 1/n\sum_{s=1}^n a_{\bar is} a_{s\bar j} \textrm{.}
\end{equation}

We get for the probability $P_{\bar i\bar i}$ of the elementary extension of fourth type (Fig.\ \ref{fig:fig6})
\begin{equation}
\label{eq:ACP1.10} P_{\bar i\bar i}= 1/(2n)\sum_{s=1}^n a_{\bar is} a_{s\bar i} \textrm{.}
\end{equation}

\section{THE ALGORITHM FOR CALCULATING THE MATRIX OF AMPLITUDES \label{ACM}}

We can calculate the probability of any elementary extension if we can calculate the matrix of amplitudes for every connected d-graph. Consider an iterative procedure for this matrix. This procedure starts from $\mathcal{G}_1$. This is the x-structure. We have
\begin{equation}
\label{eq:ACM1.1}
\begin{array}{cccc}
(\mathbf{a}(\mathcal{G}_1)) &=&\left( \begin{array}{cc} 1/2& 1/2 \\ 1/2 & 1/2 \end{array} \right) &\textrm{.}
\end{array}
\end{equation}

By \cite[Teorem~2]{1008.5169} we can get every connected d-graph from the x-structure by a sequence of elementary extensions of the considered four types. Consider the transformations of the matrix of amplitudes for each type of elementary extension. Denote by $\mathcal{G}_N$ the initial d-graph and by $\mathcal{G}_{N+1}$ the received d-graph, where $N$ is the number of x-structures in $\mathcal{G}_N$. $n$ is the width of $\mathcal{G}_N$.

\begin{figure}
	\centering	
		\includegraphics[width=4cm,trim=8cm 16cm 8cm 5cm]{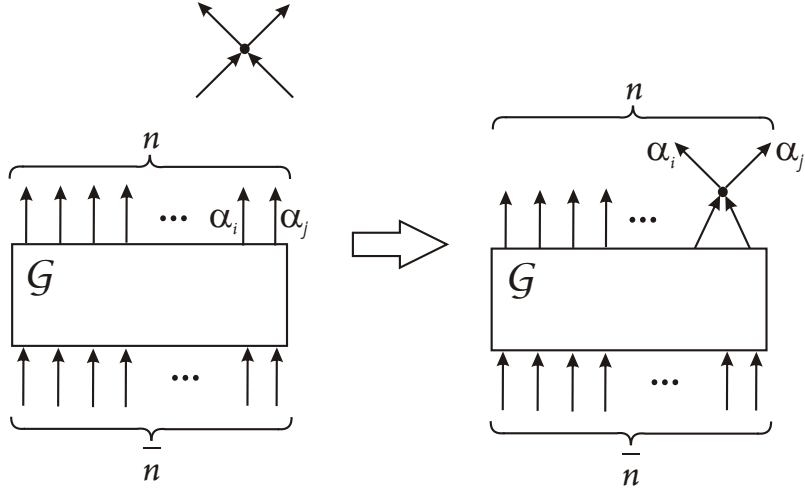}
	\caption{First type of elementary extension.}
	\label{fig:fig9}
\end{figure}

First type is an internal elementary extension to the future (Fig.\ \ref{fig:fig9}).
Two maximal monads $\alpha_i$ and $\alpha_j$ are included in the new edges. These monads are excluded from the set of maximal monads. We get two free numbers of maximal monads: $i$ and $j$. Two new maximal monads appear. Number these new maximal monads by $i$ and $j$. New maximal monads $\alpha_i$ and $\alpha_j$ are included in the same paths. These paths are all paths in which the old maximal monads $\alpha_i$ and $\alpha_j$ are included. These paths pass through one new x-structure. Then we must multiply by $1/2$. We get for the elements of  columns numbers $i$ and $j$ of $(\mathbf{a}(\mathcal{G}_{N+1}))$ 
\begin{equation}
\label{eq:ACM1.2}
a_{\bar ri}(\mathcal{G}_{N+1})=a_{\bar rj}(\mathcal{G}_{N+1})=1/2(a_{\bar ri}(\mathcal{G}_N)+ a_{\bar rj}(\mathcal{G}_N))\textrm{,}
\end{equation}
where $i$ and $j$ are fixed, and $\bar r$ ranges from $\bar 1$ to $\bar n$. Other columns and the size of the matrix of amplitudes are not changed.

\begin{figure}
	\centering	
		\includegraphics[width=4cm,trim=8cm 16cm 8cm 5cm]{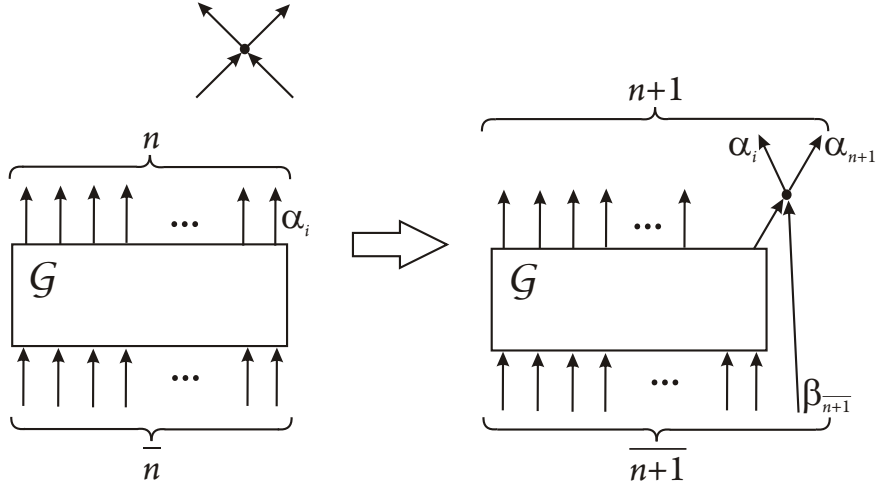}
	\caption{Second type of elementary extension.}
	\label{fig:fig10}
\end{figure}

Second type is an external elementary extension to the future (Fig.\ \ref{fig:fig10}).
One maximal monad $\alpha_i$ is included in the new edge. This monad is excluded from the set of maximal monads. We get $i$ as free number of maximal monads. Two new maximal monads and one new minimal monad appear. Number these new maximal monads by $i$ and $n+1$, and new minimal monad by $\overline{n+1}$. New maximal monad $\alpha_i$ is included in the same paths as the old maximal monad $\alpha_i$. These paths pass through one new x-structure. Then we must multiply by $1/2$. We get for the elements of the column number $i$ of $(\mathbf{a}(\mathcal{G}_{N+1}))$  
\begin{equation}
\label{eq:ACM1.3}
a_{\bar ri}(\mathcal{G}_{N+1})=(1/2)a_{\bar ri}(\mathcal{G}_N) \textrm{,}
\end{equation}
where $i$ is fixed, and $\bar r$ ranges from $\bar 1$ to $\bar n$. New maximal monad $\alpha_{n+1}$ is included in the same paths as new maximal monad $\alpha_i$. We get new column number $n+1$ with the following elements.
\begin{equation}
\label{eq:ACM1.4}
a_{\bar r(n+1)}(\mathcal{G}_{N+1})=a_{\bar ri}(\mathcal{G}_{N+1}) \textrm{,}
\end{equation}
where $i$ is fixed, and $\bar r$ ranges from $\bar 1$ to $\bar n$. The new minimal monad $\beta_{\overline{n+1}}$ is connected by directed paths only with the maximal monads $\alpha_i$ and $\alpha_{n+1}$. Each connection includes one path that passes through one x-structure. We get new row number $\overline{n+1}$ with the following elements.
\begin{equation}
\label{eq:ACM1.5}
a_{(\overline{n+1}) i}(\mathcal{G}_{N+1})= a_{(\overline{n+1}) (n+1)}(\mathcal{G}_{N+1})=1/2 \textrm{,}
\end{equation}
where $i$ is fixed.
\begin{equation}
\label{eq:ACM1.6}
a_{(\overline{n+1}) r}(\mathcal{G}_{N+1})=0 \textrm{,}
\end{equation}
where $r$ ranges from 1 to $i-1$ and from $i+1$ to $n$. The size of the matrix of amplitudes is increased by 1 from $n$ to $n+1$.

\begin{figure}
	\centering	
		\includegraphics[width=4cm,trim=8cm 14cm 8cm 7cm]{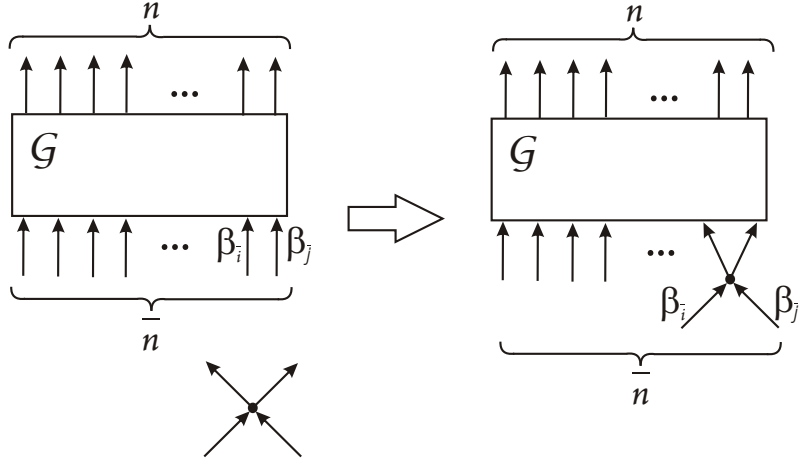}
	\caption{Third type of elementary extension.}
	\label{fig:fig11}
\end{figure}

Third type is an internal elementary extension to the past (Fig.\ \ref{fig:fig11}).
Two minimal monads $\beta_{\bar i}$ and $\beta_{\bar j}$ are included in the new edges. These monads are excluded from the set of minimal monads. We get two free numbers of minimal monads: $\bar i$ and $\bar j$. Two new minimal monads appear. Number these new minximal monads by $\bar i$ and $\bar j$. New minimal monads $\beta_{\bar i}$ and $\beta_{\bar j}$ are included in the same paths. These paths are all paths in which the old minimal monads $\beta_{\bar i}$ and $\beta_{\bar j}$ are included. These paths pass through one new x-structure. Then we must multiply by $1/2$. We get for the elements of rows numbers $\bar i$ and $\bar j$ of $(\mathbf{a}(\mathcal{G}_{N+1}))$ 
\begin{equation}
\label{eq:ACM1.7}
a_{\bar ir}(\mathcal{G}_{N+1})=a_{\bar jr}(\mathcal{G}_{N+1})=1/2(a_{\bar ir}(\mathcal{G}_N)+ a_{\bar jr}(\mathcal{G}_N))\textrm{,}
\end{equation}
where $\bar i$ and $\bar j$ are fixed, and $r$ ranges from 1 to $n$. Other rows and the size of the matrix of amplitudes are not changed.

\begin{figure}
	\centering	
		\includegraphics[width=4cm,trim=8cm 14cm 8cm 7cm]{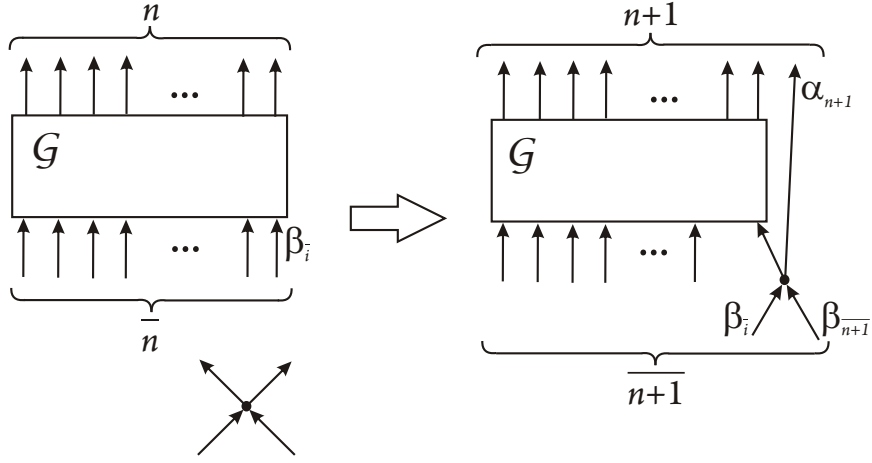}
	\caption{Fourth type of elementary extension.}
	\label{fig:fig12}
\end{figure}

Fourth type is an external elementary extension to the past (Fig.\ \ref{fig:fig12}).
One minimal monad $\beta_{\bar i}$ is included in the new edge. This monad is excluded from the set of minimal monads. We get $\bar i$ as free number of minimal monads. Two new minimal monads and one new maximal monad appear. Number these new minimal monads by $\bar i$ and $\overline{n+1}$, and new maximal monad by $n+1$. New minimal monad $\beta_{\bar i}$ is included in the same paths as the old minimal monad $\beta_{\bar i}$. These paths pass through one new x-structure. Then we must multiply by $1/2$. We get for the elements of the row number $\bar i$ of $(\mathbf{a}(\mathcal{G}_{N+1}))$  
\begin{equation}
\label{eq:ACM1.8}
a_{\bar ir}(\mathcal{G}_{N+1})=(1/2)a_{\bar ir}(\mathcal{G}_N) \textrm{,}
\end{equation}
where $\bar i$ is fixed, and $r$ ranges from 1 to $n$. New minimal monad $\beta_{\overline{n+1}}$ is included in the same paths as new minimal monad $\beta_{\bar i}$. We get new row number $\overline{n+1}$ with the following elements.
\begin{equation}
\label{eq:ACM1.9}
a_{(\overline{n+1})r}(\mathcal{G}_{N+1})=a_{\bar ir}(\mathcal{G}_{N+1}) \textrm{,}
\end{equation}
where $\bar i$ is fixed, and $r$ ranges from 1 to $n$. The new maximal monad $\alpha_{n+1}$ is connected by directed paths only with the minimal monads $\beta_{\bar i}$ and $\beta_{\overline{n+1}}$. Each connection includes one path that passes through one x-structure. We get new column number $n+1$ with the following elements.
\begin{equation}
\label{eq:ACM1.10}
a_{\bar i (n+1)}(\mathcal{G}_{N+1})= a_{(\overline{n+1}) (n+1)}(\mathcal{G}_{N+1})=1/2 \textrm{,}
\end{equation}
where $i$ is fixed.
\begin{equation}
\label{eq:ACM1.11}
a_{\bar r(n+1)}(\mathcal{G}_{N+1})=0 \textrm{,}
\end{equation}
where $\bar r$ ranges from $\bar 1$ to $\overline{i-1}$ and from $\overline{i+1}$ to $\bar n$. The size of the matrix of amplitudes is increased by 1 from $n$ to $n+1$.

We can calculate the probability of any elementary extension of any finite connected d-graph by a finite number of steps of this algorithm.

\section{DISCUSSION \label{DIS}}

In this model, the order relation of a d-graph will be causal in the dynamical sense, and not only in name. The probability of elementary extension to the future depends only on the past sets of maximal monads that take part in this extension. Similarly, the probability of elementary extension to the past depends only on the future sets of minimal monads that take part in this extension. Only the normalization constant depends on other parts of a d-graph. It is inversely proportional to the width of a d-graph.

\begin{figure}
	\centering	
		\includegraphics[width=4cm,trim=8cm 17cm 8cm 7cm]{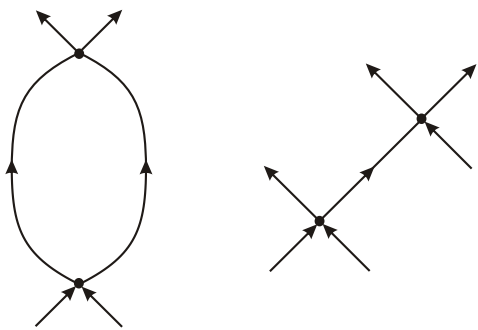}
	\caption{The d-graphs that consist of two x-structure.}
	\label{fig:fig13}
\end{figure}

The numbering of monads has not physical meaning. If d-graphs are differed only by the numbering of monads, they are identical. For example, there are only two d-graphs that consist of two x-structure (Fig.\ \ref{fig:fig13}).

The main hypothesis of this study is that cyclic structures describe particles. The goal is to identify certain structures and real kind of particles. D. Finkelstein proposed such identification \cite[Fig. 4]{STC4}. But the models are different and these structures cannot be described by a d-graph. We must consider other structures. The properties of particles are considered now as manifestations of symmetry. Consequently we must investigate the symmetry of d-graphs. The most important is the symmetry of the matrix of amplitudes because the dynamics depends on this symmetry. For example, the matrix of amplitudes is not changed if we replace any vertex by a double edge. A double edge is figured on the left of Fig.\ \ref{fig:fig13}. But we cannot exclude the consideration of double edges because such exclusion violates the normalization of probabilities. The properties of particles must correspond to the symmetry of some blocs of matrixes of amplitudes.

The important problem is the correspondence between complex amplitudes in quantum theory and amplitudes of causal connections in this model. In this concept of the dynamics, we assume that complex amplitudes in quantum theory are not fundamental quantities \cite{0904.1862}. This is a mathematical tool for the approximated description of d-graphs by quantum fields in continuous spacetime.

A sequential growth of a d-graph is a sequential growth of directed paths. The amplitudes of causal connections linearly depend on the directed paths (\ref{eq:BA1.1}). Then we have the linear dynamical equations for amplitudes of causal connections (\ref{eq:ACM1.2})-(\ref{eq:ACM1.11}). Probabilities depend on the particular type of path (Fig.\ \ref{fig:fig8}). This path forms a loop with a new x-structure. Such path is a product of two directed paths. Loops are ``a square'' of directed paths. By assumption, probabilities linearly depend on loops. Then we have the quadratic equations for probabilities (\ref{eq:ACP1.1})-(\ref{eq:ACP1.2}).

In quantum theory, we have linear dynamical equations for amplitudes and quadratic equations for probabilities too. But the quantum amplitudes are complex numbers. A complex phase describes some internal cyclic processes. But the quantum theory does not describe the internal structure of these processes. The goal of the considered model is to describe these internal structures as some structures of d-graphs. Matrixes are useful for a description of such structures. For example, these may be an adjacency matrix or an incidence matrix. It is possible that complex numbers in quantum amplitudes are a representation of some real matrix algebra. For example, there is the following isomorphism that takes each complex number to the $2\times 2$ real matrix.
\begin{equation}
\label{eq:DIS1.1}
\begin{array}{cccccccc}
1 & \to & \left( \begin{array}{cc} 1& 0 \\ 0 & 1 \end{array} \right) &\textrm{,}&
-1 & \to & \left( \begin{array}{cc} -1& 0 \\ 0 & -1 \end{array} \right) &\textrm{,}
\end{array}
\end{equation}
\begin{equation}
\label{eq:DIS1.2}
\begin{array}{cccccccc}
i & \to & \left( \begin{array}{cc} 0& 1 \\ -1 & 0 \end{array} \right) &\textrm{,}&
-i & \to & \left( \begin{array}{cc} 0& -1 \\ 1 & 0 \end{array} \right) &\textrm{.}
\end{array}
\end{equation}
In this matrix representation, the complex conjugation is the change of the sign of the matrixes (\ref{eq:DIS1.2}). In quantum theory, the complex conjugation is the time reversal. The direction of time is the direction of causal connections. In this paper, we consider only the matrixes with positive elements. These matrixes do not describe the direction of causal connections. We can define some matrix that describes this direction. The element number $ij$ of this matrix describes the direction of causal connection between monads $\gamma_i$ and $\gamma_j$. It is positive if $\gamma_i\prec \gamma_j$, and it is negative if $\gamma_j\prec \gamma_i$.

The other important problem is the background of spacetime. A pregeometry must describe all: spacetime and matter. Usually in causal set theory, discrete elements are uniformly distributed in spacetime \cite{BMSorkin}. In this case, we can describe empty spacetime but we must add matter ad hoc. A pregeometry must form nonuniform hierarchical structures. Loops can be such structures. In this case, the correspondence between spacetime and pregeometry can be more complicated \cite{1006.2320}. 

This model is useful for numerical simulation. This, however, is still work for a future.

\section*{ACKNOWLEDGEMENTS}
I am grateful to several colleagues for extensive discussions on this subject, especially Alexandr V. Kaganov and Vladimir V. Kassandrov.

\end{document}